\documentclass[a4paper,amsmath,amssymb,floatfix,prb,footinbib,twocolumn,superscriptaddress,preprintnumbers]{revtex4}

\usepackage{graphicx}    
\usepackage{dcolumn}     
\usepackage{bm}          
\usepackage{color}
\usepackage[colorlinks=true ,citecolor=blue]{hyperref}
\usepackage{dsfont}

\date{\today}

\begin{document}

\title{Adiabatic pumping in the quasi-one-dimensional triangle lattice}

\author{Michael Schulze}
\affiliation{Physikalisches Institut, Albert-Ludwigs-Universit\"at, D-79104 Freiburg, Germany}
\affiliation{Freiburg Institute for Advanced Studies, Albert-Ludwigs-Universit\"at, D-79104 Freiburg, Germany}
\author{Dario Bercioux}
\email{dario.bercioux@frias.uni-freiburg.de}
\affiliation{Freiburg Institute for Advanced Studies, Albert-Ludwigs-Universit\"at, D-79104 Freiburg, Germany}
\author{Daniel F. Urban}
\affiliation{Physikalisches Institut, Albert-Ludwigs-Universit\"at, D-79104 Freiburg, Germany}
\affiliation{Fraunhofer Institute for Mechanics of Materials IWM, D-79108 Freiburg, Germany}

\begin{abstract}
We analyze the properties of the quasi-one-dimensional triangle lattice emphasizing the occurrence of flat bands and 
band touching via the tuning of the lattice hopping parameters and on-site energies. The spectral properties of the 
infinite system will be compared with the transmission through a finite piece of the lattice with attached semi-infinite leads.
Furthermore, we investigate the adiabatic pumping properties of such a system:
depending on the transmission through the lattice, this results in nonzero integer charge transfers or transfers that increase linearly with the lattice size.
\end{abstract}

\maketitle

\section{Introduction}

Quantum pumping is a process, where a periodic variation of system parameters leads to a density flux of particles through the system in spite of the absence of an external voltage.
The first proposal for quantum pumping was formulated by Thouless in 1983,\cite{thouless}
and it considered particles in a periodic system described by Bloch wave-functions.
The Thouless approach is restricted to insulating systems and here the number of particles transferred in one period is always quantized to integer values.

Another pumping approach is based on elastic scattering matrices. It was formulated by Brouwer\cite{brouwer} and B\"uttiker\cite{buttiker,buttiker_alternativ,buttiker_long} 
and allowed to explain experimental results by Switkes \emph{et al.}\cite{switkes} for finite open systems.
In general, the particle transfer within a pumping cycle is found not to be quantized to integer values.
Both approaches are equivalent if transmission through the finite static system is suppressed along the whole pumping cycle\cite{graf:2008,math_compare} and in this case, 
the charge transfer is quantized.

In this Article, we investigate quantum pumping in a quasi-one-dimensional lattice, \emph{i.e.} a one-dimensional lattice model with a basis 
of two (or several, in general) lattice sites. These kinds of systems are basic model systems for the verification of fundamental 
phenomena such as the Aharonov-Bohm effect in presence of electron-interaction\cite{vidal:2000,platero:2010}, the Aharonov-Casher effect\cite{bercioux:2004}, 
or the combination of the two effects.\cite{bercioux:2005} In some recent experiments, the Aharonov-Casher effect has been observed in quasi-one-dimensional 
chains realized with topological insulators.\cite{qu:2011}
Quasi-one-dimensional linear chains have also been proposed for studying effects of spin-polarization in presence of spin-orbit interaction\cite{Aharony:2008}
and for studying the difference between boson and fermion dynamics in a cold-atoms experiment.\cite{flat_fermion_boson} 

\begin{figure}[ht]
\begin{center}
    \includegraphics[width=\columnwidth]{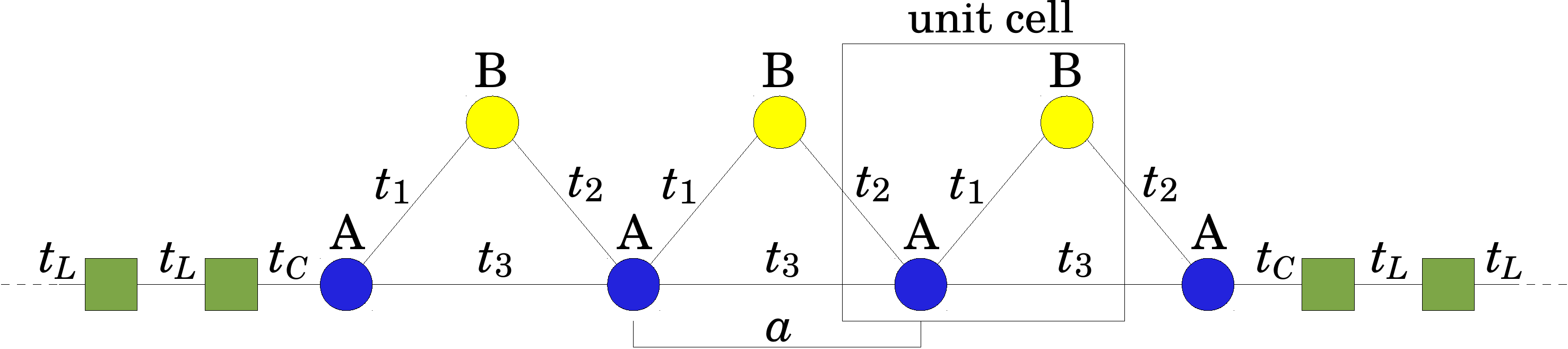}
    \caption{Tight-binding triangle lattice with semi-infinite leads.
    A number of unit cells ($N$) with the two sites A and B are connected with hopping parameters
    $t_1$, $t_2$ and $t_3$ to form  the triangle lattice of length $L=N\:a$. Note that an additional
    $(N+1)$th A--site is required to complete the finite triangle chain.
    \label{fig:triangle_finite}}
\end{center}
\end{figure}

Here, we will focus on the quasi-one-dimensional triangle lattice (Fig.~\ref{fig:triangle_finite}). This lattice is characterized by a basis of two lattice sites. 
Its band structure exhibits flat bands and band touching as a function of the lattice parameters. First we work out the conditions for the appearance of these 
peculiar features. Then we relate the properties of the infinite lattice to the transmission probability of a finite piece of lattice which is connected via 
semi-infinite leads to some particle reservoirs. Finally, we apply Brouwer's formalism\cite{brouwer} for adiabatic quantum pumping (AQP) to the scattering matrix 
of this finite-size structure.
We find that the charge transfer is finite for a pumping parameter cycle surrounding a band touching configuration, even if along the cycle the  transmission 
is inhibited. Further, if the pumping cycle traverses configurations with nonzero transmission, the charge transfer has on average a linear dependency on $N$, 
the number of unit-cells of the triangle lattice (see Fig.~\ref{fig:triangle_finite}). However, this result 
is valid only for the charge transfer but not for the pumping current. In fact, in order to fulfill the adiabatic approximation
which is underlying the derivation of Brouwer's formula, the pumping frequency has 
to decrease with the number of unit-cells.\cite{note:fabio}

%
%
%
%

\section{The triangle lattice}

\subsection{Spectral properties}
%
%
\begin{figure}[ht]
\begin{center}
    \includegraphics[width=\columnwidth]{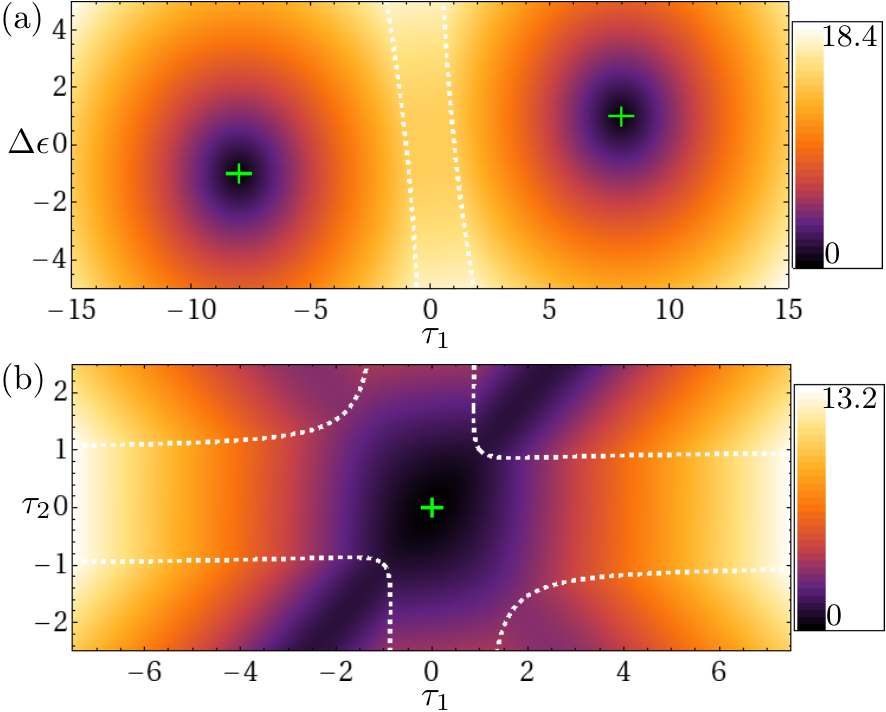}
    \caption{(Color online) Illustration of the band gap for the infinite triangle lattice.
    Brightness visualizes the value of the energy gap. White dashed curves correspond to the appearance of a flat band.
    The crosses (green) are located at band-touching (a) or band-crossing points (b).
    Corresponding parameters in (a): $\tau_2 = 8$; (b): $\Delta \epsilon = 0.5$.\label{fig:gap}}
\end{center}
\end{figure}
%
%

We define the quasi-one-dimensional triangle lattice within the tight-binding formalism. It has a basis containing two lattice sites  
A and B with on-site energies $\epsilon_\text{A}$ and $\epsilon_\text{B}$ and hopping parameters $t_1$, $t_2$ and $t_3$ as sketched 
in Fig.~\ref{fig:triangle_finite}. For the infinite lattice | no leads | the energy spectrum of the system can be obtained via Bloch's 
theorem. The Hamiltonian in reciprocal space  reads
%
%
\begin{align}
\label{triangle_hamilton_reci}
    \mathcal{H}(\kappa) =
    \begin{pmatrix}
    \epsilon_\text{A} + 2 t_3 \cos \kappa & t_1+t_2 \text{e}^{- \text{i} \kappa}\\
    t_1+t_2 \text{e}^{\text{i} \kappa} & \epsilon_\text{B}
    \end{pmatrix},
\end{align}
%
%
with $\kappa=k \: a$. The eigenvalues can be expressed with dimensionless variables as
%
%
\begin{subequations}\label{general_triangle_spec}
\begin{align}
    \label{en:one}
    \epsilon_{1,2}(\kappa) & = \cos \kappa \pm \Big(\left[\Delta\epsilon+ \tau_1 \tau_2 + 
    \cos\kappa\right]^2 + \tau_1^2 + \tau_2^2  \nonumber \\
    & \hspace{1.5cm}-\tau_1^2 \tau_2^2 -2 \tau_1 \tau_2\Delta\epsilon \Big)^\frac{1}{2} 
\\	
    \label{en:two}
    & = \cos \kappa \pm\Big(\!\left[ \Delta \epsilon + \cos \kappa \right]^2 \!
    + \tau_1^2 + \tau_2^2 +2 \tau_1 \tau_2 \cos \kappa\!\Big)^\frac{1}{2}
    \!\! , 
\end{align}
\end{subequations}
%
%
where all parameters are expressed in units of $t_3$:  $\tau_i = t_i/t_3$, $\epsilon(\kappa) = (E(\kappa) - (\epsilon_\text{A}+\epsilon_\text{B})/2)/t_3$, 
$\Delta \epsilon =  (\epsilon_\text{A}-\epsilon_\text{B})/2t_3$. We can also introduce the dimensionless coupling constant $c=t_c^2/(t_Lt_3)$  between the leads and the central system. 
In the following we will focus on the regime of intermediate coupling, $c=1$.
Since we consider the system in absence of external magnetic fields, all lattice parameters are real-valued due to time-reversal-symmetry  
[$\mathcal{H}^*(\kappa)=\mathcal{H}(\kappa)$]. However, we assume the possibility of sign changes of the hopping parameters and on-site-energies. 
Depending on the parameters, the energy gap between the two bands, which is defined as the difference between the minimum of the upper and the maximum of the 
lower band, is visualized via the color scale in Fig.~\ref{fig:gap}(a)--(b). Here, the spectrum \eqref{general_triangle_spec} exhibits a one-dimensional 
Dirac point [Fig.~\ref{fig:spec}(a)], band crossing with flat bands, and opening of gaps as visualized in Fig.~\ref{fig:spec}(e) and \ref{fig:spec}(g), respectively. 
In order to observe a band touching or crossing the square root in \eqref{general_triangle_spec} must vanish for a specific $\kappa$. This is realized for either of the 
three configurations
%
%
\begin{subequations} \label{triangle_closing}
\begin{alignat}{2}
  &\tau_1 = \tau_2 & \Delta \epsilon &= 1 \\
  &\tau_1 = -\tau_2 & \Delta \epsilon &= -1 \\
  &\tau_1 = \tau_2 = 0 &\hspace{2cm} |\Delta \epsilon| &\le 1 \text{ .} \label{triangle_closing_flat}
\end{alignat}
\end{subequations}
%
%
The first two cases are included in Fig.~\ref{fig:gap}(a). On the other hand, case \eqref{triangle_closing_flat} corresponds to the splitting of the lattice 
into a linear chain (A-sites) with a cosine band and the isolated B-sites leading to a flat band (see Fig.~\ref{fig:spec}(e) and Fig.~\ref{fig:gap}(b)).
The parameter $\Delta \epsilon$ is then responsible for the relative position between the flat band and the cosine band. If $|\Delta \epsilon| < 1$, the 
bands cross at two points within the first Brillouin zone.

The general condition for the occurrence of a flat band is immediately obtained by \eqref{en:one} and reads
%
%
\begin{equation}
  \tau_1^2 + \tau_2^2 - \tau_1^2 \tau_2^2 -2 \tau_1 \tau_2  \Delta \epsilon = 0.
\end{equation}
%
%
It is visualized in Fig.~\ref{fig:gap}(a)--(b) by the white dashed lines.

\subsection{Transport properties}
%
%
\begin{figure}[b]
\begin{center}
  \includegraphics[width=\columnwidth]{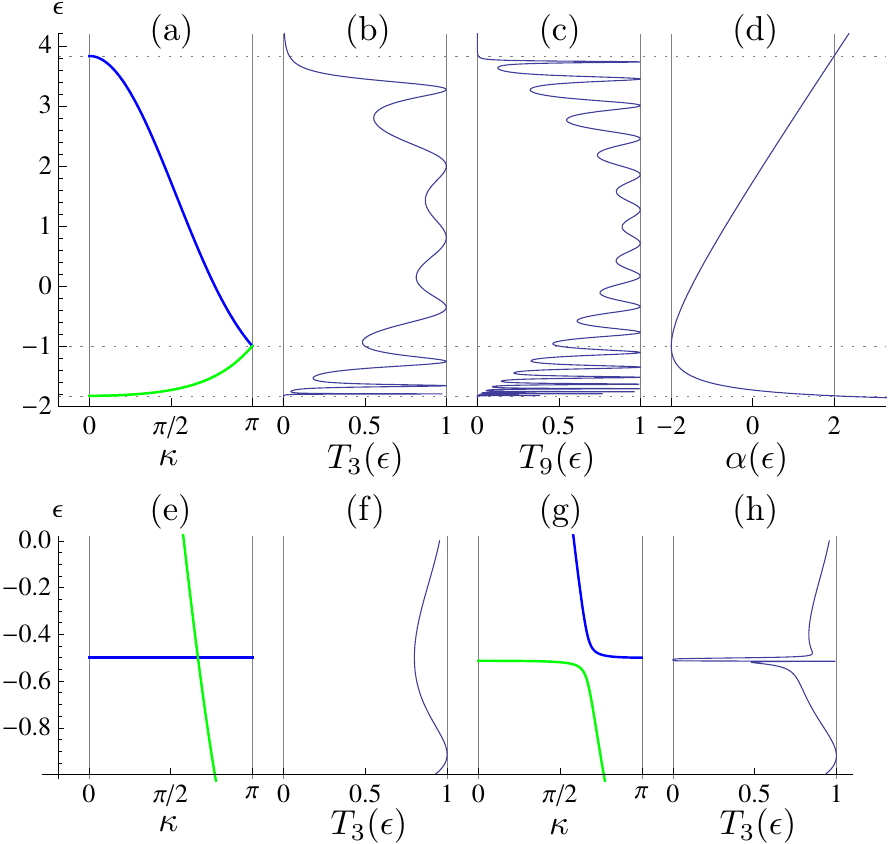}
  \caption{Collection of energy bands $\epsilon(\kappa)$ of the infinite lattice, transmission probability $T_N(\epsilon)$ and
  $\alpha(\epsilon)$ for the finite open lattice ($N =$ number of unit-cells). (a)--(d): band touching with 
  $\tau_1=\tau_2=\Delta \epsilon = 1$; (e),(f): Band crossing with $\tau_1=\tau_2=0$, $\Delta \epsilon = 0.5$ ; 
  (g),(h): Splitting with $\tau_1=\tau_2=0.1$, $\Delta \epsilon = 0.5$. \label{fig:spec}}
\end{center}
\end{figure}
%

We consider a finite piece of the triangle lattice coupled to particle reservoirs via two semi-infinite one-channel leads, which are modeled via linear chains. 
The calculation of the elastic scattering matrix ($\mathcal{S}$-matrix) can be done with the help of the 
Fisher-Lee-relation\cite{fisher_lee,sols,fisher_lee_derivation_tight_binding,datta} which 
relates the transmission and reflection amplitudes contained in the $\mathcal{S}$-matrix to the retarded equilibrium Green's functions of the system. 
Both quantities, the $\mathcal{S}$-matrix and the 
retarded Green's function, contain the information about the solution of the underlying Schr\"odinger equation describing the quantum system. 
While the time-dependent Green's function represents the full-system 
time evolution, the scattering matrix relates the probability amplitudes of the outgoing to the incoming states in the asymptotic region. 
The relation between both quantities was first found by Fisher 
and Lee \cite{fisher_lee}. More general derivations were later done for continuous\cite{sols} and tight-binding systems.\cite{fisher_lee_derivation_tight_binding} 
The elements of the Green's function of the infinite system including the leads, which are required in order to determine the transport properties, 
are determined through the calculation of the self-energies of the two semi-infinite leads which are then added to the finite lattice Hamiltonian.\cite{datta}
The problem is then reduced to a finite set of algebraic equations. The details of the calculation of the scattering matrix are presented in App.~\ref{smatrix}. 
The analytic expression for the resulting $\mathcal{S}$-matrix reads
%
%
\begin{subequations} \label{triangle_s_matrix}
\begin{align}
  \textbf{S}_{1 2} &= \textbf{S}_{2 1} = 
  \frac{-2 \text{i} \hspace{0.05cm} \sigma}{\textbf{B}_{1 1} +( \beta_1 + \beta_2) \textbf{B}_{1 2} - \beta_1 \beta_2  \textbf{B}_{2 2}}, \\
  \textbf{S}_{j j} &=  \frac{ 2 \text{i} \hspace{0.05cm} \sigma \hspace{0.05cm} (\textbf{B}_{1 2} - \beta_j \textbf{B}_{2 2})}
  {\textbf{B}_{1 1} +( \beta_1 + \beta_2) \textbf{B}_{1 2} - \beta_1 \beta_2  \textbf{B}_{2 2}} -1,
\end{align}
\end{subequations}
%
%
where $\textbf{S}_{i j}$ is the reflection/transmission amplitude for an incoming wave in lead $j$ to an outgoing wave in lead $i$. 
Here the indices $\{1,2\}$ label the left and the right lead, respectively. The Matrix $\textbf{B}$ is defined as
%
%
\begin{equation}
  \textbf{B} = \begin{pmatrix}
  0 & 1 \\
  -1 & \alpha
\end{pmatrix}^{N-1} \text{ .}
\end{equation}
%
%
Here, $N$ is the number of unit cells, c.f. Fig.~\ref{fig:triangle_finite}. The variables $\alpha$, $\beta_1$, $\beta_2$ and $\sigma$ depend on the tight-binding parameters and
are given by
%
%
\begin{equation} \label{triangle_variables}
\begin{aligned}
  \sigma =&\frac{\epsilon  + \Delta \epsilon }{\epsilon+   \Delta \epsilon + \tau_1 \tau_2 },\\
  \alpha =& \frac{\epsilon^2- \Delta \epsilon^2-\tau_1^2 - \tau_2^2}{\epsilon + \Delta \epsilon + \tau_1 \tau_2}, \\
  \beta_j =&  \frac{\epsilon^2- \Delta \epsilon^2-\tau_j^2 }{\epsilon + \Delta \epsilon + \tau_1 \tau_2} +\text{i} \hspace{0.05cm}  \sigma,
\end{aligned}
\end{equation}
%
%
where $\epsilon$ represents the energy of the incoming particles and is defined in units of $t_3$ identically to $\epsilon(\kappa)$ 
in \eqref{general_triangle_spec}. The wide-band-limit (energy independent lead properties) is used here for the $\mathcal{S}$-matrix. 
The matrix-elements of $\textbf{B}$ essentially control whether the lattice is a conductor or an insulator, because it is the quantity 
that mainly controls the denominator of $\textbf{S}_{1 2}$ (and therefore the transmission probability $T=|\textbf{S}_{1 2}|^2$) and 
contains the $N$-dependency. The parameter $\alpha$ determines whether the elements of $\textbf{B}$ grow exponentially with $N$ or oscillate.
If the definition of $\alpha$ in \eqref{triangle_variables} is solved for $\epsilon$, the spectrum of the infinite lattice \eqref{en:two} 
is recast with $\alpha$ replacing $2 \cos \kappa$:
%
%
\begin{equation}
  \epsilon_{1,2} = \frac{\alpha}{2} \pm \bigg(\Big[ \Delta \epsilon + \frac{\alpha}{2} \Big]^2 + \tau_1^2 + \tau_2^2 + \tau_1 \tau_2 \alpha\bigg)^\frac{1}{2}.
\end{equation}
%
%
One can deduce that if and only if the value of $\epsilon$ is part of the energy spectrum, there is a real valued $\kappa$ 
satisfying $\alpha= 2 \cos \kappa$ and hence $|\alpha| \le 2$. An elementary analysis of matrix $\textbf{B}$ yields that 
its elements grow exponentially with $N$ if $|\alpha| > 2$. With $|\alpha| < 2$, the dependency is periodic and at $|\alpha| = 2$ 
it is linear. Respective explicit expressions for $\textbf{B}$ are given in Eqs.\ (\ref{eq:B:osz}) and (\ref{eq:B:decay}) in 
the Appendix. We conclude that the transmission $T=|\textbf{S}_{1 2}|^2$ through the finite lattice is suppressed exponentially 
with $N$ if the particle energy is not part of the spectrum of the infinite lattice. For energies that belong to the spectrum 
the transmission oscillates and the number of resonances is related to $N$. The transmission is visualized together with the 
parameter $\alpha$ as a function of the particle energy $\epsilon$ in Fig.~\ref{fig:spec}. For one example of energy bands 
[Fig.~\ref{fig:spec}(a)] the transmission is presented for $N=3$ [Fig.~\ref{fig:spec}(b)] and $N=9$ [Fig.~\ref{fig:spec}(c)]. 
The parameter $\alpha$, which mainly controls the transmission, is shown in Fig.~\ref{fig:spec}(d). One can see that the interval 
$\alpha\in[-2,2]$ corresponds exactly to the spectrum. Transmission outside of this interval is suppressed as a function of $N$ 
[cf. Fig.~\ref{fig:spec}(b) and \ref{fig:spec}(c)]. Figure~\ref{fig:spec}(h) illustrates the effect on the transmission when 
a gap opens in the energy spectrum, namely the suppression  of the transmission inside the gap.

\subsection{Adiabatic quantum pumping properties}
%
%
\begin{figure*}[t]
\begin{center}
  \includegraphics[width=\textwidth]{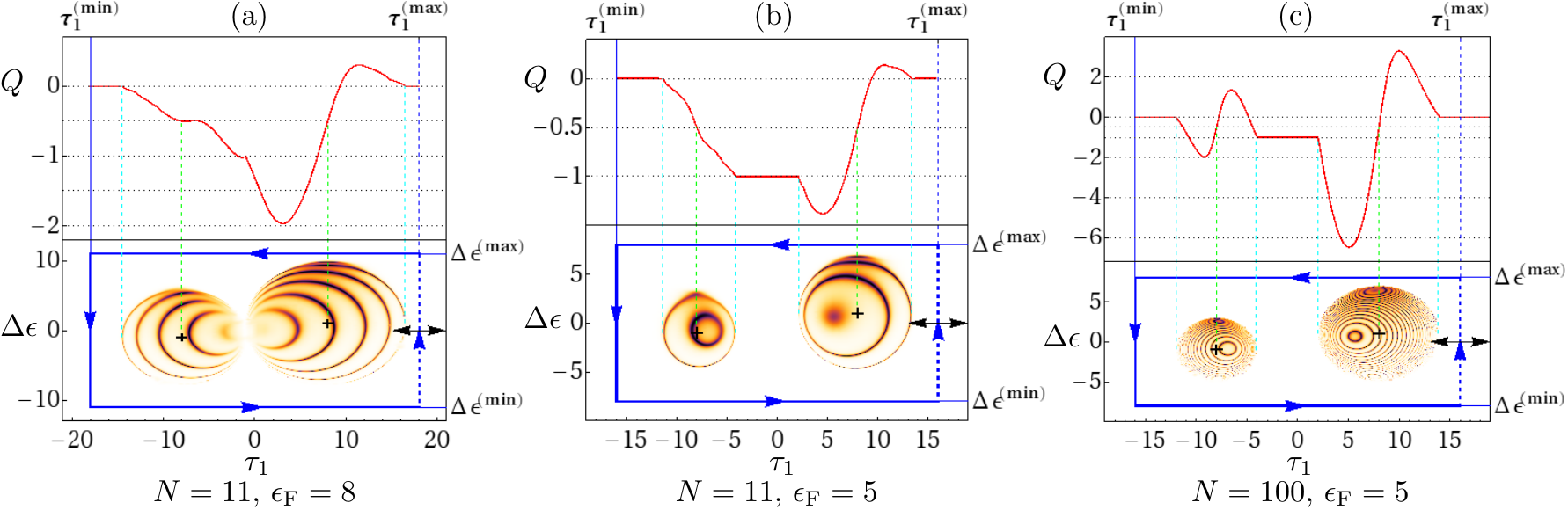}
  \caption{(Color online) Varying parameters $\tau_1$ and $\Delta \epsilon$ (path $\mathcal{A}$) around 
  two band touching points (marked with black crosses) for different Fermi-energies $\epsilon_\text{F}$ and $N$. 
  The lower panels show the parameter cycle (blue) on the density plot of transmission $T$ (black: $T = 1$, white: $T=0$). 
  In the upper panels the charge transfer is plotted versus $\tau_1^\text{(max)}$. Vertical dashed lines (cyan) indicate 
  the boundary between insulating and transmitting regions. For all panels: $\tau_2=8$.  \label{fig:tde_quant}}
\end{center}
\end{figure*}
%
%

Having determined the scattering matrix of the finite triangle lattice, the scattering approach to the AQP\cite{brouwer} can be applied. 
By varying slowly two of the system parameters, a net current can be produced in absence 
of an external bias. In the adiabatic regime the current is proportional to the frequency $\omega$ of the periodic variation. 
On the other hand, the charge transfer $Q$, i.e. the integral of the current over one parameter cycle, is independent of $\omega$. 
Brouwer's formula for the charge transfer in the zero temperature limit reads\cite{brouwer}
%
%
\begin{equation} 
  \label{brouwer_formula}
  Q_{j} = \frac{e}{\pi} \int_A \text{d}X_1 \text{d}X_2 \hspace*{2pt} \text{Im} \left\{ \text{Tr} \left[ \left(  \frac{\partial \textbf{S}}{\partial X_2} \frac{\partial \textbf{S}^{\dagger}}{\partial X_1}\right)_{j j} \right] \right\} \text{ ,}
\end{equation}
%
%
where $j$ is the lead-index and $X_1$ and $X_2$ are the pumping parameters. 
A non-zero charge transfer is the result of taking into account 
that a non-static system allows inelastic scattering processes. However, due to the adiabatic approximation, the shifts in energy 
are infinitesimally small and can be expressed via derivatives of the elastic scattering matrix.
The adiabatic approximation becomes exact in the limit $\omega \rightarrow 0$. 

The trace in Eq.\ (\ref{brouwer_formula}) is over the channels and can be omitted in the case of one-channel-leads. The scattering matrix has to be evaluated at the Fermi-energy, 
which will be expressed with the dimensionless scaled parameter $\epsilon_\text{F}$ in analogy to the scaling of $\epsilon(\kappa)$ 
in~\eqref{general_triangle_spec}. The choice of the lead $j$ determines the sign of the charge transfer. It is convenient here to 
choose $j=\text{R}$ which fixes the result to the transfer from the left to the right side in Fig.~\ref{fig:triangle_finite}.

For the system we consider here, there are three possible choices of pumping parameters $\{X_1 ,X_2\}$ in Brouwer's formula (\ref{brouwer_formula}): \{$\tau_1$,$\tau_2$\}, \{$\tau_1$,$\Delta \epsilon$\} and \{$\tau_2$,$\Delta \epsilon$\}. 
Exchanging the parameters $\tau_1$ and $\tau_2$ is equivalent to the exchange of the orientation along the lattice axes. Hence, it is equivalent 
to a sign change of the charge transfer. It also follows that the charge transfer vanishes if $\tau_1 = \tau_2$ holds for the whole cycle. 
After fixing the set of pumping parameters, we study how the choice of the parameter cycle influences the charge transfer. In the following we 
will focus on the two choices $\{\tau_1,\Delta \epsilon\}$ ($\Rightarrow$ path $\mathcal{A}$) and $\{\tau_1,\tau_2\}$ ($\Rightarrow$ path $\mathcal{B}$).
For both cases the charge transfer is plotted as a function of the extreme of the pumping parameter path and is always expressed in units of the electron charge ($e=1$).

Figure~\ref{fig:tde_quant} shows examples where path $\mathcal{A}$ is varied around band touching points 
[c.f. Fig.~\ref{fig:gap}(a) and \ref{fig:spec}(a)]. Here the lower panels show the transmission probability together with the pumping parameter 
cycles. In the upper panels the charge transfer is plotted versus the maximum value of $\tau_1$ in the cycle. 
In the following we summarize the main features of the transferred charge $\mathcal{Q}$ along paths $\mathcal{A}$ and $\mathcal{B}$:

\begin{enumerate}
\item \emph{Crossing of a spectral gap} |
If the pumping path is chosen such that it always remains within a gap of the finite system energy spectrum, the charge transfer $Q$ is quantized to an integer value as shown in 
Figs.~\ref{fig:tde_quant}(a)--\ref{fig:tde_quant}(c) and \ref{fig:tt_quant}(b), which is in agreement with Refs.\ [\onlinecite{brouwer,graf:2008,math_compare,g0,quant}]
If, on the other hand, the pumping path traverses the Dirac-like point of the energy spectrum [crosses in the lower 
panels of Fig.~\ref{fig:tde_quant}], the charge transfer $Q$ is also quantized but to an half--integer value so as observed also in graphene.~\cite{prada:2009,prada:2011}
%
\item \emph{Dependence on the number of unit cells $N$}	|
The charge transfer $Q$  on average depends linearly on the number of unit cells $N$. Figure~\ref{fig:tt_quant} illustrates this relation for a parameter path of type
$\mathcal{B}$ traversing configurations with nonzero transmission. 
The charge transfer for this cycle is calculated for different $N$ (i.e. different lengths of the lattice), and is visualized 
in Fig.~\ref{fig:tt_quant}(d).
%
%
\begin{figure}[b]
  \centering
  \includegraphics[width=\columnwidth]{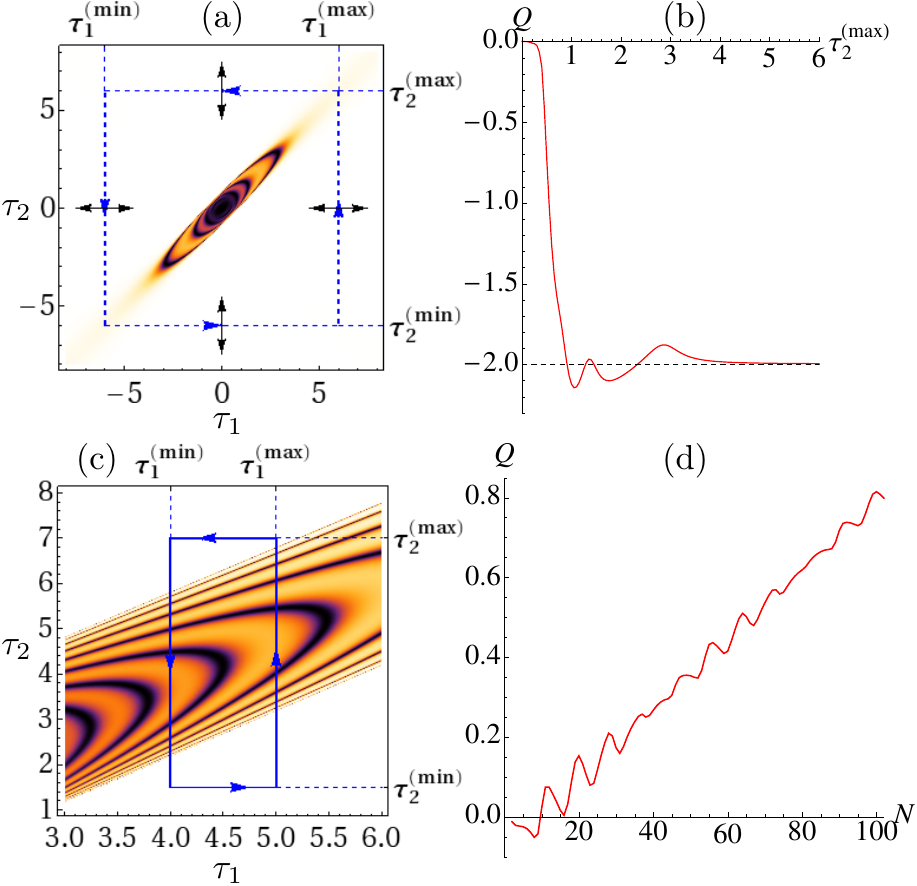}
  \caption{(Color online) Demonstration of different charge transfer results with pumping parameters $\tau_1$ and 
  $\tau_2$ (path $\mathcal{B}$). Panels (a) and (b): Symmetric pumping cycle with 
  $\tau_i^\text{(max)}=-\tau_i^\text{(min)}$, $\Delta \epsilon=0.5$,  $\epsilon_\text{F}=0$ and $N=11$ 
  around crossing point. Panels (c) and (d): Charge transfer of pumping in transmitting region increases 
  linearly with lattice length $N$. $\Delta \epsilon=1.8$, $\epsilon_\text{F}=1$. Brightness of panels 
  (a) and (c) corresponds to transmission $T$ (Black: $T = 1$, White: $T=0$). \label{fig:tt_quant}}
\end{figure}
%
%
%
\item \emph{Influence of the coupling to the leads} |
For a fixed parameter path we find that both in the case of very weak or very strong coupling $c$, the pumped charge is quantized to a constant value. 
This is because in both limits the parameter paths remain in the energy gap of the finite size system.
%
\item \emph{Crossing of a flat band} | In the case of path $\mathcal{B}$ it is possible to cross the parameter configuration corresponding to a flat band 
in the infinite lattice,\cite{note:density} 
cf. Eqs.~(\ref{triangle_closing}). In Fig.~\ref{fig:tt_quant}(a) and \ref{fig:tt_quant}(b) the parameters are varied around this flat-band configuration. 
The bands cross at $\tau_1=\tau_2=0$ if $|\Delta \epsilon| \le 1$ [see Fig.~\ref{fig:spec}(e)]. This corresponds to the B-site (flat band) completely decoupled 
from the rest of the chain (cosine band). However, if the parameter cycle surrounds this point [see Fig.~\ref{fig:tt_quant}(a)], the structure of the 
whole system | with nonzero coupling $\tau_1$ and $\tau_2$, \emph{e.g.} as in Fig.~\ref{fig:spec}(g) | is responsible for the charge transfer. 
Figure~\ref{fig:tt_quant}(b) shows that the charge transfer becomes $\mathcal{Q}= -2e$ for large pumping cycles that purely traverse parameter 
configurations where transmission is inhibited. The potential benefits of crossing a flat band is always canceled by the fact 
that a parameter path of type $\mathcal{B}$ has to cross the flat-band configuration twice.
\end{enumerate}

\section{Conclusion}

We have presented the general spectral properties of the triangular lattice. We have shown that by changing the system parameters, 
the spectrum shows a one-dimensional Dirac cone, a flat band and other interesting features. The transport properties of the finite length 
lattice strongly depend on the system parameters. Thus, within the scattering matrix approach to the adiabatic pumping a large 
range of results can be obtained. It turns out that, 
if the parameter configurations of the cycle remains in an insulating area (no transmission), the charge transfer is 
quantized to an integer value. If the parameter path traverses the Dirac-like point, the charge transfer is quantized to an half--integer value. 
At zero transmission the Fermi-energy is positioned between the bands (without touching them). In this case, the scattering matrix only contains nonzero 
reflection amplitudes that vary along the parameter cycle. Although all particles are reflected in the static case, 
the variation in time allows for scattering processes that lead to a net charge transfer $Q$ through the system that 
is independent of the length of the lattice.

If the parameter path crosses areas with nonzero transmission, the charge transfer generally takes a non-integer value. 
If transmission is allowed, the charge transfer depends sensitively on all used parameters. If particles are transmitted 
through the lattice, the amount of energy which is gained or lost on average depends linearly on $N$, the number of unit-cells of the lattice 
[c.f. Fig. \ref{fig:tt_quant}(d)]. This is related to the parameter change being applied to all parts of the lattice. 
A perturbation affecting only one specific unit cell of the lattice does not lead to a linear $N$-dependency. 

Note that this linear dependency on $N$ does not implicate the generation of an arbitrary large current with increasing 
lattice size. Because of the approximations used in this model the pumping frequency $\omega$ needs to be sufficiently 
small. This requires $\omega$ to decrease with increasing length of the quasi-one-dimensional lattice.

\acknowledgments We thank Piet Brouwer and Heinz-Peter Breuer for useful discussions. The work of MS and DB is 
supported by the Excellence Initiative of the German Federal and State Governments.

\appendix

\section{Calculation of the S-matrix}
\label{smatrix}

\noindent The Hamiltonian in position space has the form
%
%
\begin{equation}
  \textbf{H} = \begin{pmatrix}
  \textbf{H}_{\text{L}} 							& 		\textbf{T}_1 	&		0		\\
  \textbf{T}^{\dagger}_1	& 		\textbf{H}_{\text{S}} 		&		\textbf{T}^{\dagger}_2 \\
  0								& 		\textbf{T}_2 	&		\textbf{H}_{\text{L}}	
\end{pmatrix} ,
\end{equation}
%
%
where $\textbf{H}_{\text{L}}$ is the Hamiltonian of the (semi-infinite) linear chain and $\textbf{H}_{\text{S}}$ is the one of the lattice sample. 
$\textbf{T}_1 $ and $\textbf{T}_2 $ are the matrices that connect lead sites with sample sites and here have only one non-zero element.
The retarded Green's function, defined via the relation
%
%
\begin{equation}
  \label{eq:Def:GF}
  [(E + \text{i}  \eta) \mathrm{\mathbf{\mathds{1}}} - \textbf{H}] \hspace{2pt} \textbf{G}^\text{R}(E) = \mathrm{\mathbf{\mathds{1}}} \text{ ,}
\end{equation}
%
%
with an infinitesimal $\eta>0$, can also be partitioned into sub-matrices by 
%
%
\begin{equation}
  \label{eq:GF:decompose}
  \textbf{G}^\text{R} (E) = \begin{pmatrix}
  \textbf{G}_{1 1} 	&  \textbf{G}_{1 \text{S}}  &	\textbf{G}_{1 2}		\\
  \textbf{G}_{\text{S} 1}	&  \textbf{G}_{\text{S}}	 &	\textbf{G}_{\text{S} 2}  \\
  \textbf{G}_{2 1}	&  \textbf{G}_{2 \text{S}}  &	\textbf{G}_{2 2}		
\end{pmatrix}.
\end{equation}
%
%
The Fisher-Lee-relation relates these Green's function elements to the elements of the scattering matrix. \cite{fisher_lee,datta,sols,fisher_lee_derivation_tight_binding}

In our calculation, we model the leads as linear chains with dispersion relation
$E_\kappa = 2 t_3 \cos \kappa + \epsilon_\text{L}$.
It is useful to consider the wide-band-limit, where $\kappa$ is fixed and the Green's function does not depend on $E_\kappa$ nor on the on-site energy $\epsilon_\text{L}$. If
$t_3 \gg E_\kappa-\epsilon_\text{L}$,
the choice $\kappa = \frac{\pi}{2}$ ($\cos \pi/2 = 0$) is valid for all particle energies.
The Fisher-Lee-relation for this case can then be expressed as
%
%
\begin{equation} 
  \label{fisherlee}
  \textbf{S} = - \mathrm{\mathbf{\mathds{1}}} + 2 \text{i} t_3 \begin{pmatrix}
  (\textbf{G}_{\text{S}})_{1,1} & (\textbf{G}_{\text{S}})_{1,2N+1} \\
  (\textbf{G}_{\text{S}})_{2N+1,1} & (\textbf{G}_{\text{S}})_{2N+1,2N+1}
\end{pmatrix}.
\end{equation}
%
%
The elements of the Green's function are to be taken in the translational invariant region. 
Here, the outmost-sites of the triangle lattice, \emph{i.e.} A-sites $1$ and $N+1$, can be considered 
as the beginning of the translational invariant leads, cf.\ Fig. \ref{fig:triangle_finite}. These two lattice sites
correspond to the indices 1 and $2N+1$ in the matrix $\textbf{G}_{\text{S}}$. 
Therefore, effectively only four elements of the Green's function $\textbf{G}_{\text{S}}$ are needed to calculate the scattering matrix. 
In order to determine these quantities it is useful to use the sub-matrix decomposition (\ref{eq:GF:decompose}) and to extract from (\ref{eq:Def:GF}) three sub-matrix equations,
%
%
\begin{equation}
\begin{alignedat}{2}
  [(E + \text{i}  \eta) \mathrm{\mathbf{\mathds{1}}} - \textbf{H}_{\text{L}}]& \textbf{G}_{1 \text{S}} - \textbf{T}_1 \textbf{G}_{\text{S}} &&= 0, \\
  [(E + \text{i}  \eta) \mathrm{\mathbf{\mathds{1}}} - \textbf{H}_{\text{L}}]& \textbf{G}_{2 \text{S}} - \textbf{T}_2 \textbf{G}_{\text{S}} &&= 0, \\
  [(E + \text{i}  \eta) \mathrm{\mathbf{\mathds{1}}} - \textbf{H}_{\text{S}}]& \textbf{G}_{\text{S}} - \textbf{T}^{\dagger}_1 \textbf{G}_{1 \text{S}} - 
  \textbf{T}^{\dagger}_2 \textbf{G}_{2 \text{S}} &&= \mathrm{\mathbf{\mathds{1}}}.
\end{alignedat}
\end{equation}
%
%
We can solve for $\textbf{G}_{\text{S}}$ and obtain
%
%
\begin{equation}\label{green_Gs}
\begin{aligned}
  \textbf{G}_{\text{S}} &= [(E + \text{i}  \eta) \mathrm{\mathbf{\mathds{1}}} - \textbf{H}_{\text{S}} -\textbf{T}^{\dagger}_1 \mathbf g_{\text{L}} \textbf{T}_1 -\textbf{T}^{\dagger}_2 \mathbf g_{\text{L}} \textbf{T}_2 ]^{-1} \\
  &= [(E + \text{i}  \eta) \mathrm{\mathbf{\mathds{1}}} - \textbf{H}_{\text{eff}}]^{-1},
\end{aligned}
\end{equation}
%
%
where $\mathbf g_{\text{L}} = [(E + \text{i}  \eta) \mathrm{\mathbf{\mathds{1}}} - \textbf{H}_{\text{L}}]^{-1} $ is the Green's function for a single isolated lead. 
Its edge element in position space reads \cite{ferry}
%
%
\begin{equation}
  (\mathbf g_{\text{L}})_{1 1} = \frac{1}{t_3} \text{e}^{-\text{i} \kappa(E)} = -\frac{\text{i}}{t_3},
\end{equation}
%
%
where the second identity holds within the wide band limit.
The effective Hamiltonian $\textbf{H}_\text{eff}$ of the finite size lattice sample, which includes
the semi-infinite leads via their self-energies $\Sigma_{i}=\textbf{T}^{\dagger}_{i} \mathbf{g}_{\text{L}} \textbf{T}_{i}$, $i\in\{1,2\}$,
now reads 
%
%
\begin{equation*}
  (\textbf{H}_\text{eff})_{i j} = (\textbf{H}_\text{S})_{i j} - \text{i} t_3 (\delta_{i,1}\delta_{j,1} + \delta_{i,2N+1}\delta_{j,2N+1}).
\end{equation*}
%
%
In order to demonstrate the symmetry of $\textbf{H}_\text{eff}$ and to provide an appropriate calculation of 
$\textbf{G}_{\text{S}}$ the effective Hamiltonian is also divided into sub-matrices as
%
%
\begin{equation} 
  \label{triangle_effham}
  \textbf{H}_{\text{eff}} = \begin{pmatrix}
  \textbf{H}_{\ell}				& 		\textbf{T}			&		0		& \cdots		 & 0 & 0\\
  \textbf{T}^{\dagger}	& 		\textbf{H}_{\text{c}}				&		\textbf{T}	& 				 & \vdots & \vdots \\
  0				& 		\textbf{T}^{\dagger}	&		\textbf{H}_{\text{c}}		& \ddots		 & 0 & 0\\
  \vdots			&						& 	\ddots 		& \ddots 		 & \textbf{T} & 0  \\
  0				&		\cdots			& 	0			& \textbf{T}^{\dagger} & \textbf{H}_{\text{c}} 		 & \textbf{T}  \\
  0				&		\cdots			& 	0 			& 0				 & \textbf{T}^{\dagger} & \textbf{H}_{\text{r}}	
\end{pmatrix},
\end{equation}
%
%
with
%
%
\begin{align*}
  \textbf{H}_{\ell} &= 
  \begin{pmatrix}
    \epsilon_{\text{A}} -\text{i} t_3  & t_1 \\
    t_1 & \epsilon_{\text{B}}
  \end{pmatrix}, &
  \textbf{H}_{\text{c}} &= 
  \begin{pmatrix}
    \epsilon_{\text{A}}  & t_1 \\
    t_1 & \epsilon_{\text{B}}
  \end{pmatrix},\\
  \textbf{H}_{\text{r}} &= 
  \begin{pmatrix}
    \epsilon_{\text{A}} - \text{i} t_3  & 0 \\
    0 & 0
  \end{pmatrix}, &
  \textbf{T} &= 
  \begin{pmatrix}
    t_3 & 0 \\
    t_2 & 0
  \end{pmatrix}.
\end{align*}
%
%
Now we can decompose the relation
%
%
\begin{equation} 
  \label{greencalc_1}
  [(E + \text{i}  \eta) \mathrm{\mathbf{\mathds{1}}} - \textbf{H}_\text{eff}] \hspace{2pt} \textbf{G}_\text{S} = \mathrm{\mathbf{\mathds{1}}}
\end{equation}
%
%
in terms of these sub-matrices as
%
%
\begin{subequations} 
\begin{align*}
  &
  [(E+\text{i} \eta)  \mathrm{\mathbf{\mathds{1}}} -\textbf{H}_{\ell} ]  \textbf{G}_{1,j} - \textbf{T} \textbf{G}_{2,j} 
  = \delta_{j,1} \mathrm{\mathbf{\mathds{1}}}, 
  \\
  &
  [(E+\text{i} \eta)  \mathrm{\mathbf{\mathds{1}}} -\textbf{H}_{\text{c}} ] \textbf{G}_{i,j} - \textbf{T}^{\dagger} \textbf{G}_{i-1,j} - \textbf{T} \: \textbf{G}_{i+1,j} 
  = \delta_{i,j} \mathrm{\mathbf{\mathds{1}}}, 
  \\
  &
  [(E+\text{i} \eta) \mathrm{\mathbf{\mathds{1}}} -\textbf{H}_{\text{r}} ] \textbf{G}_{N+1,j} - \textbf{T}^{\dagger} \textbf{G}_{N,j} 
  = \delta_{j,N+1} \mathrm{\mathbf{\mathds{1}}}, 
\end{align*}
\end{subequations}
%
%
where the $\textbf{G}_{i,j}$ are the corresponding sub-matrices of $\textbf{G}_\text{S}$ and $ i \in [2,N]$ and $j \in [1,N+1]$ label 
the unit cells of the finite lattice sample. Note that the index $N+1$ corresponds to the outmost A-site which is treated as a fake unit-cell
having a ``disconnected'' unit-cell partner B, cf. definition of $\textbf{H}_{\text{r}}$ and Fig. \ref{fig:triangle_finite}.
The set of equations \eqref{greencalc_1} contains twelve equations for the elements of the sub-matrices which can be solved independently for every index $j$. 
Only the sub-matrices $\textbf{G}_{1,1}$,  $\textbf{G}_{1,N+1}$, $\textbf{G}_{N+1,1}$ and $\textbf{G}_{N+1,N+1}$ are relevant for the scattering matrix. 
And only the (1,1)-element (A-site) is needed from the sub-matrices. By introducing $g_i = t_3 (\textbf{G}_{i,j})_{1,1}$, the set of relevant 
algebraic equations can be simplified to read
%
%
\begin{subequations} 
\label{triangle_simple_form}
\begin{alignat}{4}
  \beta_1 & g_1 && -  g_2 && =  \delta_{j 1} \sigma, 
  \\
  \label{green_triangle_recursive} 
  \alpha & g_i && -   g_{i+1}  - g_{i-1} &&= 0, 
  \\
  \beta_2 & g_{N+1} &&- g_{N} &&=  \delta_{j,N+1}  \sigma,
\end{alignat}
\end{subequations}
%
%
where $\alpha$, $\sigma$, $\beta_1$ and $\beta_2$ are defined in Eq. (\ref{triangle_variables}).
Further, we express Eq. (\ref{green_triangle_recursive}) as
%
%
\begin{equation}
  \begin{pmatrix}
  g_{i+1} \\
  g_{i+2}
  \end{pmatrix} 
  = 
  \begin{pmatrix}
  0 & 1 \\
  -1 & \alpha
  \end{pmatrix}  
  \begin{pmatrix}
  g_{i } \\
  g_{i+1 }
  \end{pmatrix},
\end{equation}
%
%
which allows to connect the last two elements with the first ones as
%
%
\begin{align} 
\label{green_triangle_connection}
  \begin{pmatrix}
  g_{N} \\
  g_{N+1}
  \end{pmatrix} 
  = \textbf{B} 
  \begin{pmatrix}
    g_{1} \\
    g_{2}
  \end{pmatrix} , 
  && \text{with} && \textbf{B} = 
  \begin{pmatrix}
    0 & 1 \\
    -1 & \alpha
  \end{pmatrix}^{N-1}.
\end{align}
%
%
From Eqs. \eqref{triangle_simple_form} and \eqref{green_triangle_connection} it follows that
%
%
\begin{subequations} 
\label{gi_result}
\begin{align}
  g_1 &= \sigma \frac{\delta_{j 1} \left( \textbf{B}_{1 2} - \beta_1  \textbf{B}_{2 2} \right) - \delta_{j N}}{\textbf{B}_{1 1} +( \beta_1 + \beta_2) \textbf{B}_{1 2} - \beta_1 \beta_2  \textbf{B}_{2 2}}, \\
  g_N &= \sigma \frac{\delta_{j N} \left( \textbf{B}_{1 2} - \beta_2  \textbf{B}_{2 2} \right) - \delta_{j 1}}{\textbf{B}_{1 1} +( \beta_1 + \beta_2) \textbf{B}_{1 2} - \beta_1 \beta_2  \textbf{B}_{2 2}},
\end{align}
\end{subequations}
%
%
which yields the scattering matrix \eqref{triangle_s_matrix}.
The matrix $\textbf{B}$ can be evaluated analytically to read
\begin{eqnarray}
\label{eq:B:osz}
    \textbf{B} =
    \frac{1}{\sin\nu}
    \begin{pmatrix}
    -\sin((N-2)\nu) & \sin((N-1)\nu) \\
    -\sin((N-1)\nu) & \sin(N\nu)
    \end{pmatrix}
\end{eqnarray}
with $\cos\nu =\alpha/2$ for the case $|\alpha|\le2$. On the other hand, for $|\alpha|>2$ we obtain
\begin{eqnarray}
\label{eq:B:decay}
    \textbf{B} =
    \frac{1}{\sinh\nu}
    \begin{pmatrix}
    -\sinh((N-2)\nu) & \sinh((N-1)\nu) \\
    -\sinh((N-1)\nu) & \sinh(N\nu)
    \end{pmatrix} \hspace{16pt}
\end{eqnarray}
with $\cosh\nu =|\alpha/2|$.


\begin{thebibliography}{99}
\bibitem{note:density} For the infinite lattice, a flat band gives rise to an infinite density of states. 
However, for a finite-length lattice the broadening of the energy levels corresponding to the flat band results in a finite density of states.
\end{thebibliography}
\end{document}